\documentclass[%
 reprint,
superscriptaddress,
 amsmath,amssymb,
 aps,pre,hyphens,floatfix
]{revtex4-2}
\usepackage{amsfonts}
\usepackage{amsmath}
\usepackage[]{amsmath}
\usepackage{amssymb}
\usepackage{multirow} 
\usepackage{enumerate}
\usepackage[usenames,dvipsnames,svgnames,table]{xcolor}
\usepackage{IEEEtrantools}
\usepackage{graphicx}
\usepackage[normalem]{ulem}  

\begin{document}

\preprint{APS/123-QED}

\title{Neural network learning of multi-scale and discrete temporal features \\in directed percolation}
\author{Feng Gao}
\affiliation{Key Laboratory of Quark and Lepton Physics (MOE) and Institute of Particle Physics, Central China Normal University, Wuhan 430079, China}

\author{Jianmin Shen}
\affiliation{College of engineering and technology, Baoshan University, Baoshan 678000, China}
\affiliation{Key Laboratory of Quark and Lepton Physics (MOE) and Institute of Particle Physics, Central China Normal University, Wuhan 430079, China}

\author{Shanshan Wang}
\affiliation{Key Laboratory of Quark and Lepton Physics (MOE) and Institute of Particle Physics, Central China Normal University, Wuhan 430079, China}

\author{Wei Li}
\email[]{liw@mail.ccnu.edu.cn}
\affiliation{Key Laboratory of Quark and Lepton Physics (MOE) and Institute of Particle Physics, Central China Normal University, Wuhan 430079, China}
\affiliation{École Supérieure d'Informatique Électronique Automatique, Ivry-sur-Seine 94200, France}

\author{Dian Xu}
\affiliation{Key Laboratory of Quark and Lepton Physics (MOE) and Institute of Particle Physics, Central China Normal University, Wuhan 430079, China}
\date{\today}

\begin{abstract}
Neural network methods are increasingly applied to solve phase transition problems, particularly in identifying critical points in non-equilibrium phase transitions, offering more convenience compared to traditional methods. In this paper, we analyze the (1+1)-dimensional and (2+1)-dimensional directed percolation models using an autoencoder network. We demonstrate that single-step configurations after reaching steady state can replace traditional full configurations for learning purposes. This approach significantly reduces data size and accelerates training time.Furthermore, we introduce a multi-input branch autoencoder network to extract shared features from systems of different sizes. The neural network is capable of learning results from finite-size scaling. By modifying the network input to include configurations at discrete time steps, the network can also capture temporal information, enabling dynamic analysis of non-equilibrium phase boundaries. Our proposed method allows for high-precision identification of critical points using both spatial and temporal features.

\end{abstract}

\maketitle

\section{Introduction}
\label{intro}

Identifying critical points in physical systems has long been a fundamental problem in statistical physics~\cite{cardy1980directed,domany1981directed,kadanoff1990scaling,wilson1971renormalization,berges2002non,hertz2018quantum,batterman2005critical,halperin2019theory}. Traditional numerical methods~\cite{janke2008monte,aizenman1986critical} often struggle with complex systems, particularly those with large scales and high spatial dimensions. To overcome these challenges, machine learning models~\cite{chen2023study,chung2023deep,miyajima2023machine,giordana2000phase} have become increasingly valuable in identifying phase transitions.

Machine learning models are computational systems that learn patterns from data to make predictions or decisions without explicit task-specific instructions~\cite{wang2024supervised,tuo2024supervised,ma2023phase,zhang2020interpreting,jo2021absorbing}. In physics, both supervised and unsupervised learning methods are commonly used. Supervised learning~
\cite{van2017learning,canabarro2019unveiling} involves learning from labeled data, allowing the discovery of new physical laws, such as critical points and phase boundaries, in the absence of prior knowledge. Unsupervised learning~\cite{wang2016discovering,kaming2021unsupervised}, on the other hand, seeks hidden patterns or structures in unlabeled data, typically used for clustering or dimensionality reduction. Machine learning enables us to bypass complex, sometimes unverified, physical laws, providing a powerful tool for system prediction.Traditional methods, such as Monte Carlo simulations and mean-field analysis, have provided comprehensive results for equilibrium phase transitions. However, research on nonequilibrium phase transitions\cite{lubeck2004universal,menyhard1995non} remains limited.

Neural networks, in particular, have proven effective for identifying phase transitions in various statistical physics models, including the Ising model~\cite{zhang2022machine,cossu2019machine,morningstar2018deep}, potts model~\cite{chen2023study,tirelli2022unsupervised,li2018applications}, XY model~\cite{zhang2019machine,wang2017machine} and percolation~\cite{cheng2021machine}. Supervised learning frameworks such as CNNs and FCNs~\cite{alzubaidi2021review,venderley2018machine}, along with unsupervised methods like PCA and AE~\cite{wetzel2017unsupervised}, have been employed in these tasks. Additionally, approaches like siamese neural network~\cite{shen2024detecting} compare system configurations' similarities to detect critical points, while recurrent neural networks~\cite{van2018learning} capture time-dependent correlations and common features of the system. However, most existing models focus on identifying critical points within a single scale.

In existing research~\cite{shen2021supervised,shen2022machine,shen2022transfer} on the DP model, CNNs and autoencoders (AE) have been applied by directly inputting configurations at different probabilities to identify critical points. However, this approach encounters challenges when dealing with larger systems or higher dimensions, as the network parameters become too large to train effectively. Additionally, these neural network frameworks are limited to learning information from a single scale.

In this paper, we propose a multi-input branch autoencoder network that extracts common features across multiple scales, enabling the identification of critical points in systems. This approach replaces traditional finite-size scaling methods. The model encodes configurations at multiple scales, fuses the features, and then uses the decoder to reconstruct the data. By training the network to optimize its parameters, the model efficiently extracts latent features, ultimately pinpointing the system’s critical point.

The structure of this paper is as follows: First, a brief introduction to the directed percolation model is provided. Then, an autoencoder network is used to verify that single-step configurations in both the (1+1)-dimensional and (2+1)-dimensional models can replace full configurations for learning and identifying the critical point. Lastly, a multi-input branch autoencoder network is proposed to extract different types of information, enabling the identification of the critical point. The approach is validated through application to the DP model.

\section{model}

The directed percolation model~\cite{cardy1980directed,berges2002non} is widely used to investigate phase transitions between survival and extinction in diffusion processes. In this model, each lattice site is either occupied or unoccupied, and the system can be in either an active state or an absorbing state, depending on the diffusion probability. The DP process is a non-equilibrium phenomenon where the transition between active and inactive phases is continuous and exhibits universal critical behavior.

The DP model is often employed to describe percolation in heterogeneous media, such as porous rocks, where adjacent pores are connected by channels with varying permeability. In geology, the DP model is used to study the extent to which water can permeate rock formations. It is typically defined as a geometric model of connectivity in random media.

This paper examines the (1+1)-dimensional and (2+1)-dimensional versions of the DP model. In the (1+1)-dimensional case, space is represented along the horizontal axis, and time is represented along the vertical axis. Each lattice site can either be occupied (\(s_i = 1\)) or unoccupied (\(s_i = 0\)). The initial state is a fully occupied particle source, which then diffuses downward. The configuration at time \(t+1\) is determined based on the following rules:

\begin{equation}
    s_i(t+1)=
    \begin{cases}
    1 & \text{if } s_{i-1}(t)=1 \text{ and } z_{i}^{-}<p, \\
    1 & \text{if } s_{i-1}(t)=1 \text{ and } z_{i}^{-}<p, \\
    0 & \text{otherwise.}
    \end{cases}
    \label{equ1}
\end{equation}

Here, \(z_{i}^{\pm}\) represents random numbers. In the key DP model on a triangular lattice, particles can only diffuse left or right from their current position. The DP model can be viewed as a particle reaction-diffusion process, where interactions and diffusion at active and inactive sites occur according to the following four processes:

\begin{equation}
    \begin{cases}
    \text{Self-destruction:} & A \to \emptyset, \\
    \text{Diffusion:} & \emptyset + A \to A + \emptyset, \\
    \text{Offspring production:} & A \to 2A, \\
    \text{Coagulation:} & 2A \to A.
    \end{cases}
    \label{equ2}
\end{equation}

In this paper, the initial state is a fully occupied lattice. According to the diffusion rules, particles spread downward, and the system's configuration evolves over time. The following describes the system’s behavior under varying bond probabilities.

\begin{figure}[!thb]
\centering
    \includegraphics[width=0.48\textwidth]{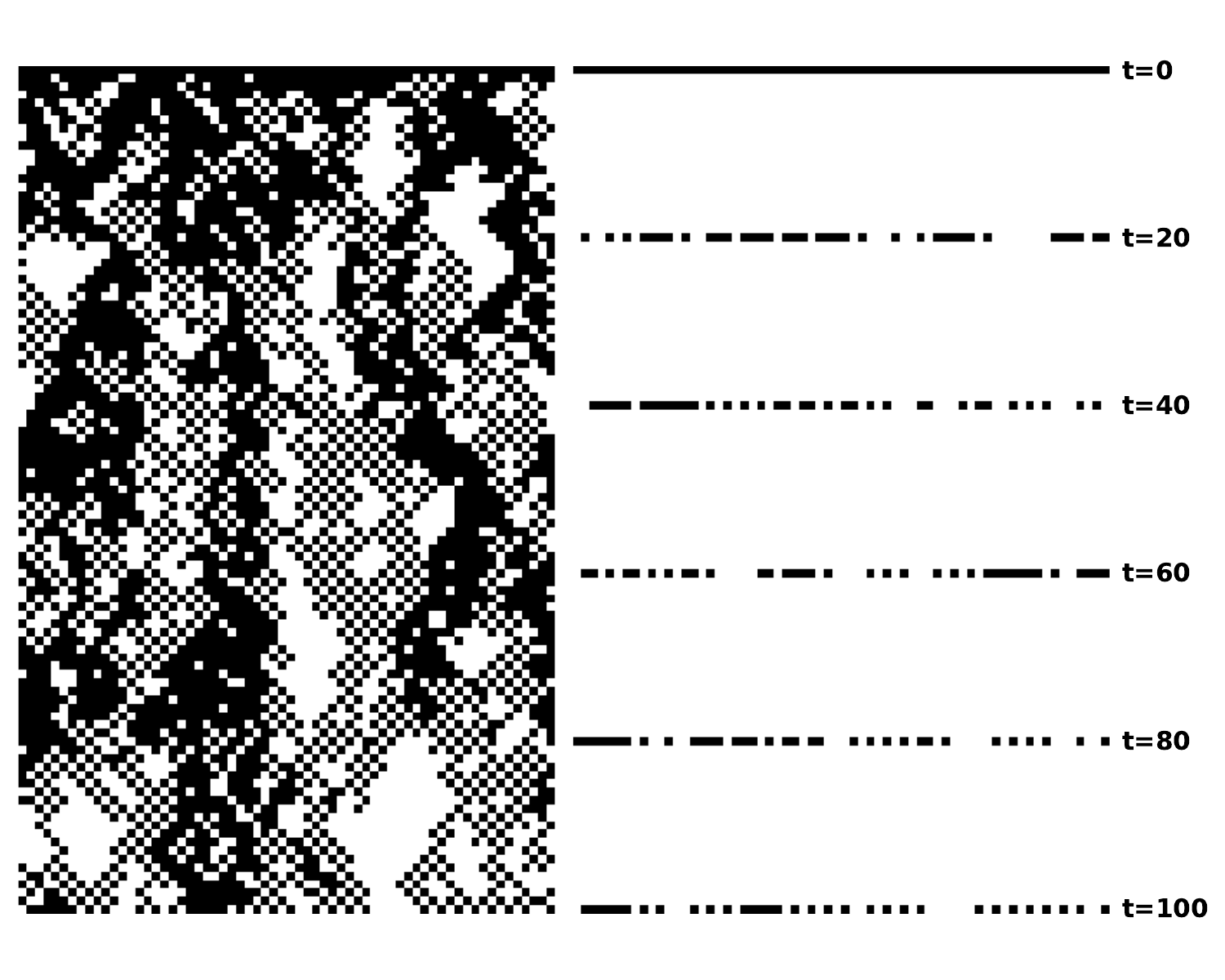} 
\caption{Configurations at different time steps for a system of size $L = 64$. \text Left: The configurations of the 1D DP model from \( t = 0 \) to \( t = 100 \). Right: The configurations at specific time steps \( t = 0, 20, 40, 60, 80, \) and \( 100 \).}
\label{configuration}
\end{figure}


\begin{figure*}[htbp]
\centering
\begin{tabular}{cccc}   
    \includegraphics[width=0.45\textwidth]{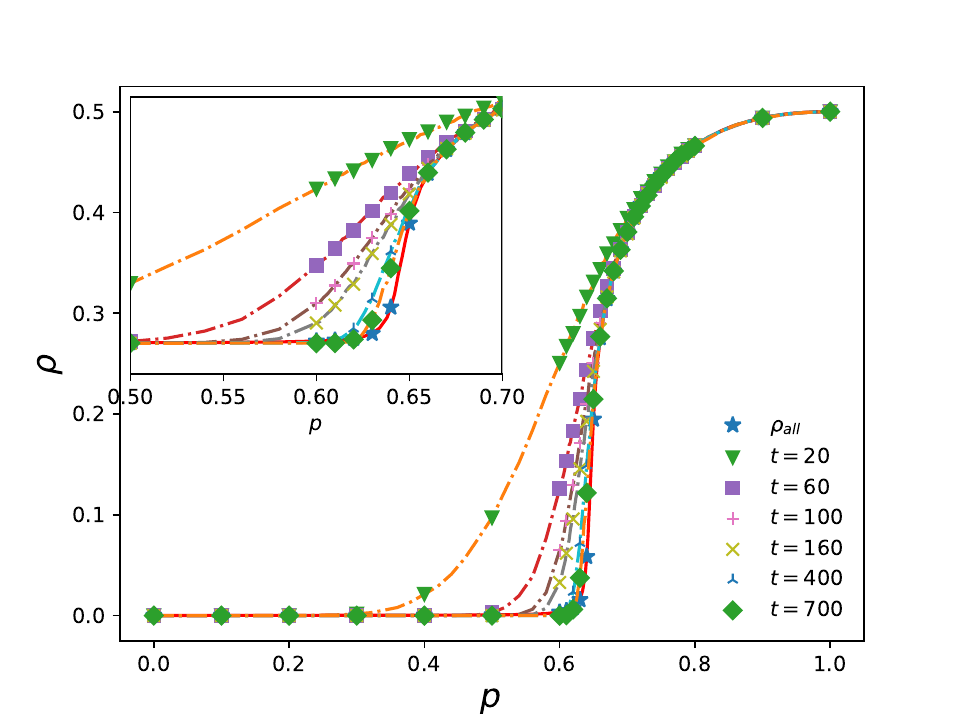} &
    \includegraphics[width=0.41\textwidth]{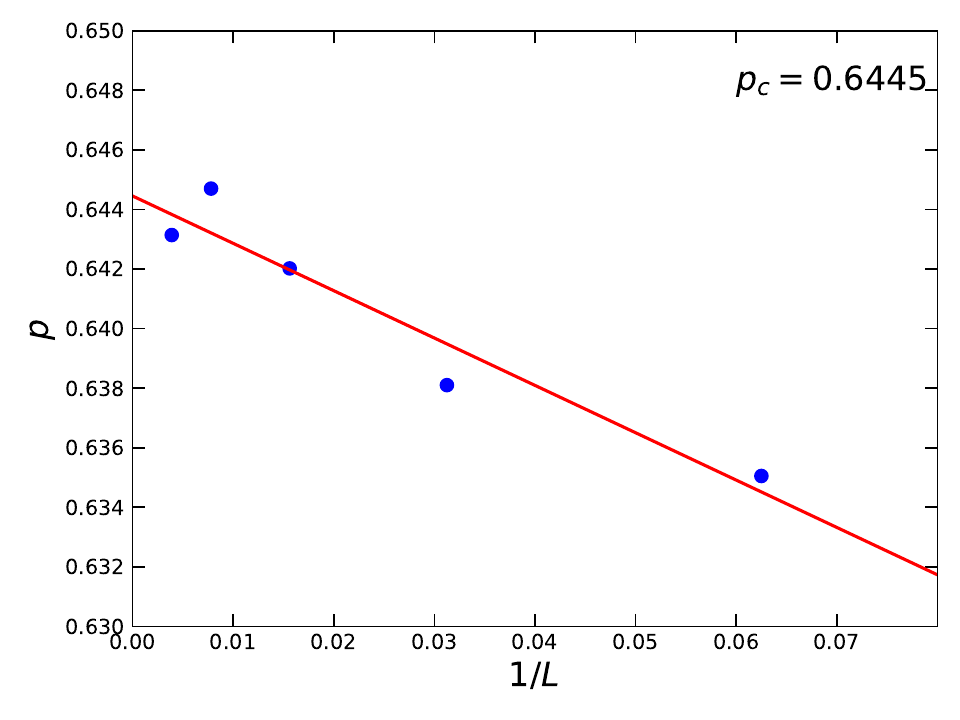} &\\
    (a) &  (b)\\
    \includegraphics[width=0.45\textwidth]{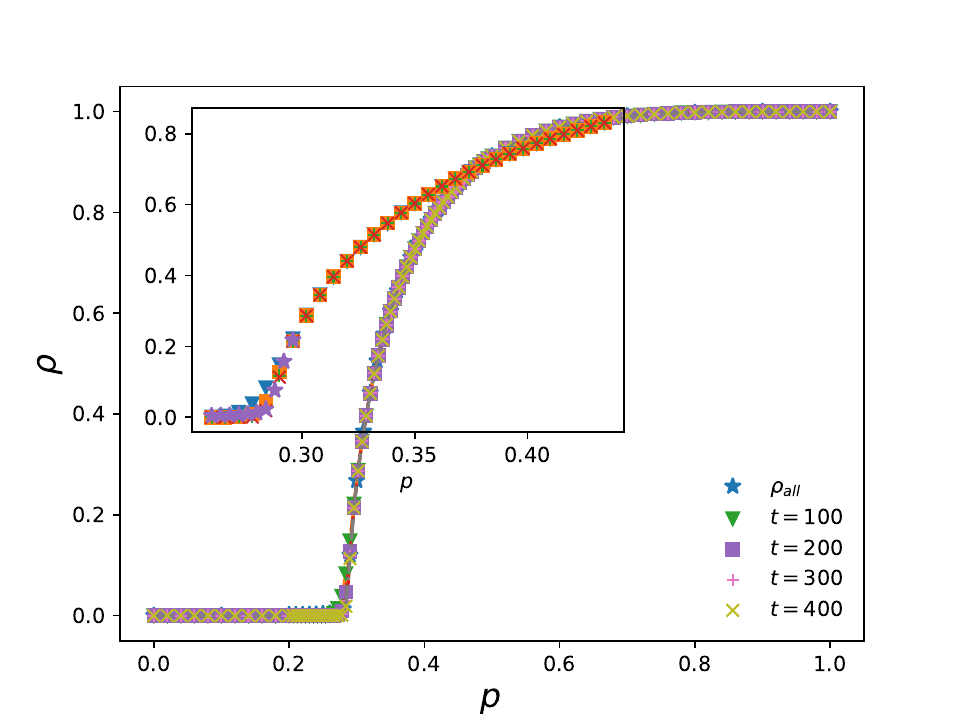} &
    \includegraphics[width=0.41\textwidth]{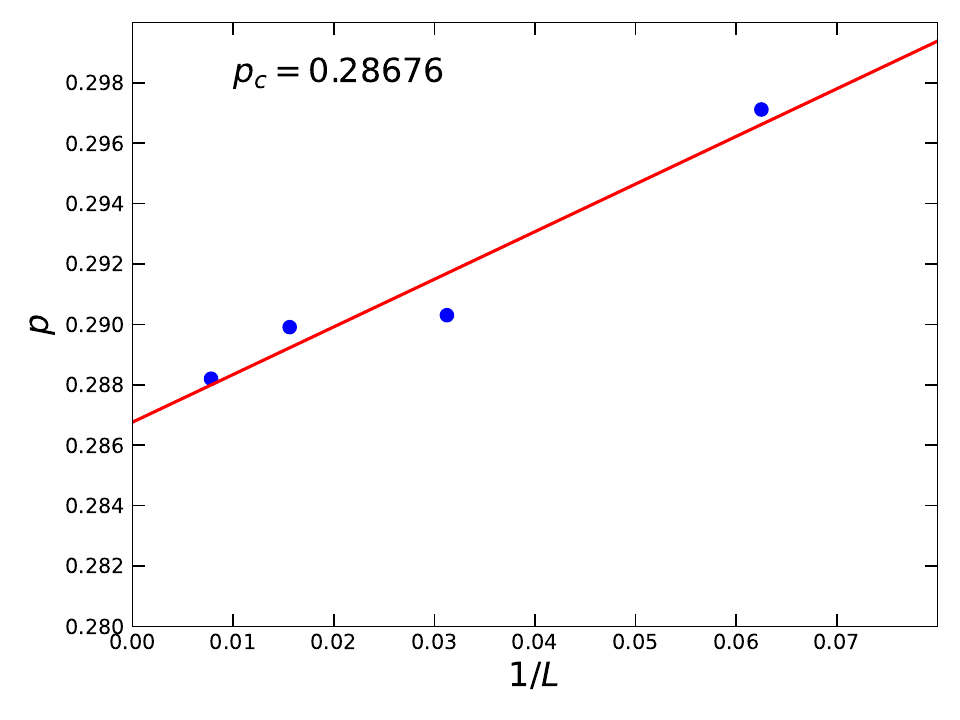} &\\
     (c)&  (d)\\
\end{tabular}
\caption{The (1+1)-dimensional DP (top) and (2+1)-dimensional DP (bottom) are shown. \textbf{(a)} and \textbf{(c)} display the average particle number density as a function of probability $p$ at different single time steps. \textbf{(b)} and \textbf{(d)} show the finite-size scaling plots of the critical probability $p_c$}
\label{mc_ans}
\end{figure*}
When \( p < p_c \), the number of particles at active sites decreases exponentially until the system reaches the absorbing state. For \( p > p_c \), the particle density saturates at a constant value, and the system remains in the active state. At the critical point, where \( p = p_c \), the particle density decays slowly, following a power-law distribution. The order parameter of the diffusion process is the density of active sites, which changes according to a power-law at the critical point:

\begin{equation}
    \rho(t) = \left\langle \frac{1}{N} \sum_i s_i(t) \right\rangle, \quad \rho^{\text{stat}} \sim (p - p_c)^\beta.
    \label{equ3}
\end{equation}

Here, \( \beta \) is the critical exponent associated with particle density. In the (1+1)-dimensional case, \( \beta \approx 0.277 \), indicating a significant change in \( \rho^{\text{stat}} \) near the transition. In the (2+1)-dimensional case, \( \beta \approx 0.58 \). Furthermore, the diffusion process exhibits a characteristic correlation length. Unlike equilibrium models, non-equilibrium critical phenomena incorporate time as an additional dimension. Non-equilibrium phase transitions are typically characterized by two independent correlation lengths: the spatial correlation length \( \xi_\perp \) and the temporal correlation length \( \xi_k \), both of which follow power-law behavior near the transition:

\begin{equation}
    \xi_\perp \sim |p - p_c|^{-\nu_\perp}, \quad \xi_k \sim |p - p_c|^{-\nu_k}.
    \label{equ4}
\end{equation}

The critical exponents \( \nu_\perp \) and \( \nu_k \) govern these length scales. For infinite systems, starting with a fully occupied lattice, the particle density \( \rho(t) \) follows:

\begin{equation}
    \rho(t) \sim t^{-\alpha}f \left( \Delta t \right)^{1/\nu_k},
    \label{equ5}
\end{equation}

where \( \alpha = \beta/\nu_k \) is the critical exponent, and \( f \) is a function with the same functional form across all DP models. To quantitatively analyze spatial and temporal equilibrium, techniques like lattice models are essential. Recent advances, such as Monte Carlo simulations and neural networks, have become valuable tools for identifying critical points in these systems.

\section{Neural Network based Approach}

Machine learning has become increasingly important for identifying phase transitions, particularly when model details are unknown. In such cases, unsupervised algorithms, such as autoencoders, are highly effective. This paper employs an autoencoder network consisting of an encoder and a decoder. The encoder compresses high-dimensional input data into low-dimensional representations (latent variables or features), aiming to retain key information while minimizing data loss. The decoder reconstructs the data from these low-dimensional features, verifying whether the encoder has captured sufficient information.

\begin{figure}[!thb]
\centering
    \includegraphics[width=0.48\textwidth,trim=0 100 330 220,clip]{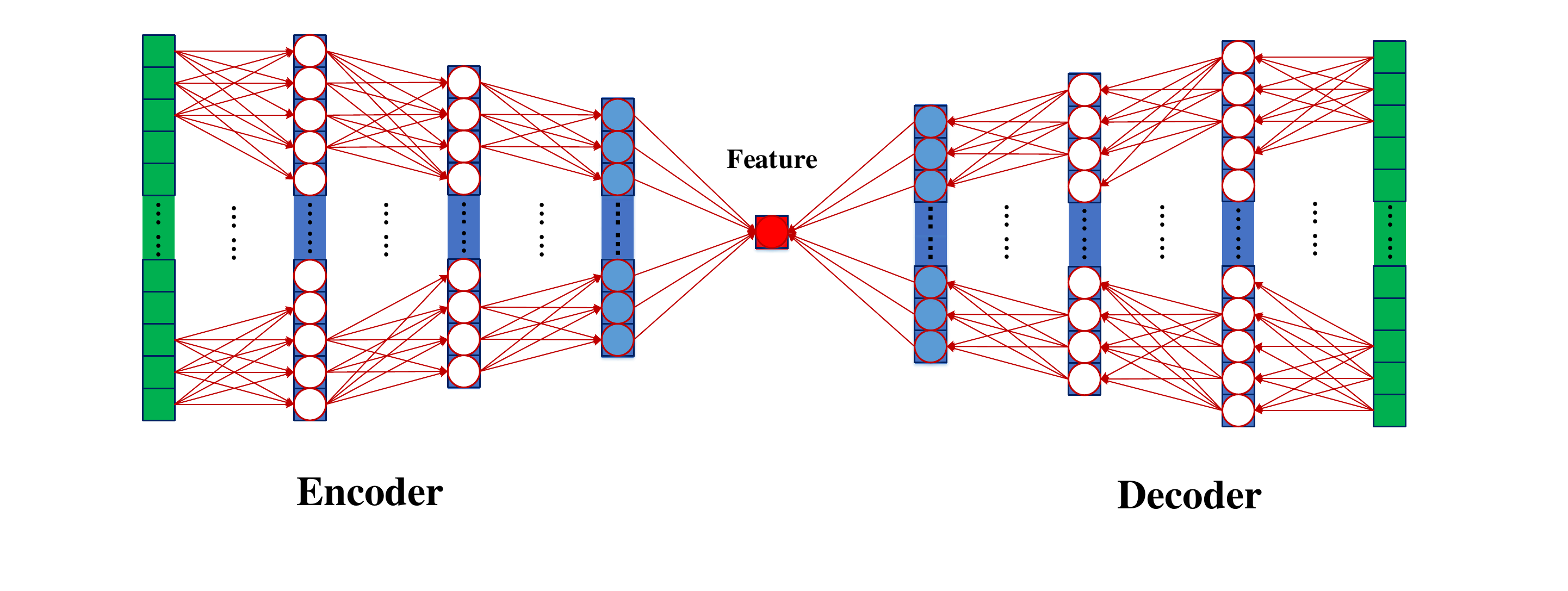} 
    \caption{Neural network schematic structure of autoencoder.}
    \label{ae}
\end{figure}

The network architecture is shown in Fig~\ref{ae}, the encoder processes the input data, which is flattened from high-dimensional configurations, through three fully connected layers to extract a single latent feature that represents the input DP configuration. The decoder structure is symmetric to that of the encoder. Since the input data consists of DP model configurations generated by Monte Carlo simulations, configurations at different times are used. For larger systems, this approach reduces the complexity of training by using single-step configurations instead of full-time configurations.

\begin{figure*}[htbp]
\centering
\begin{tabular}{cccc}   
    \includegraphics[width=0.45\textwidth]{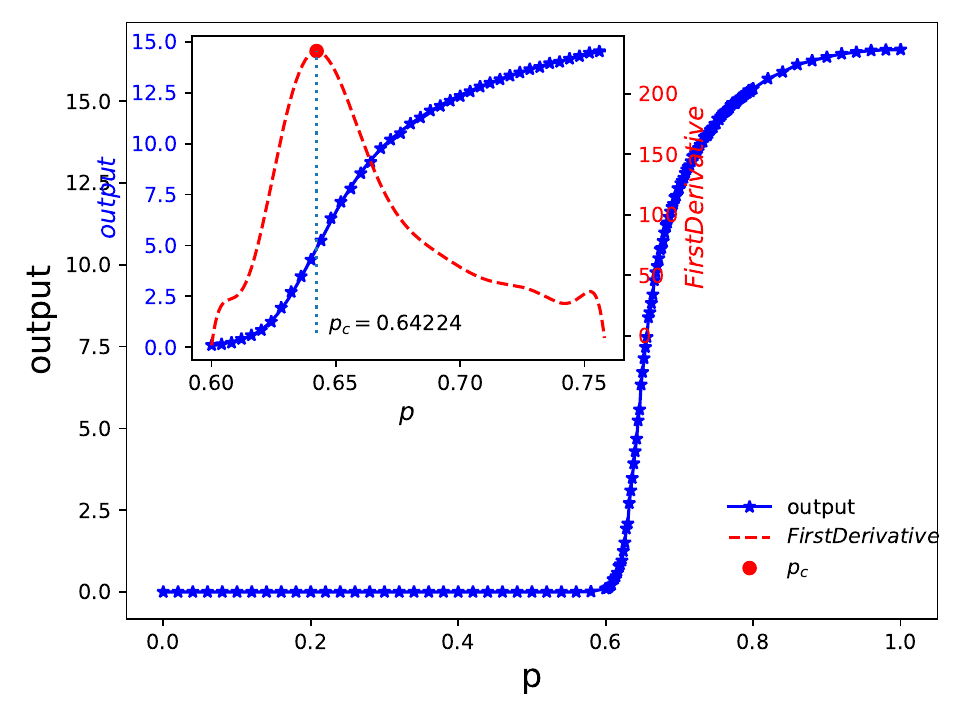} &
    \includegraphics[width=0.45\textwidth]{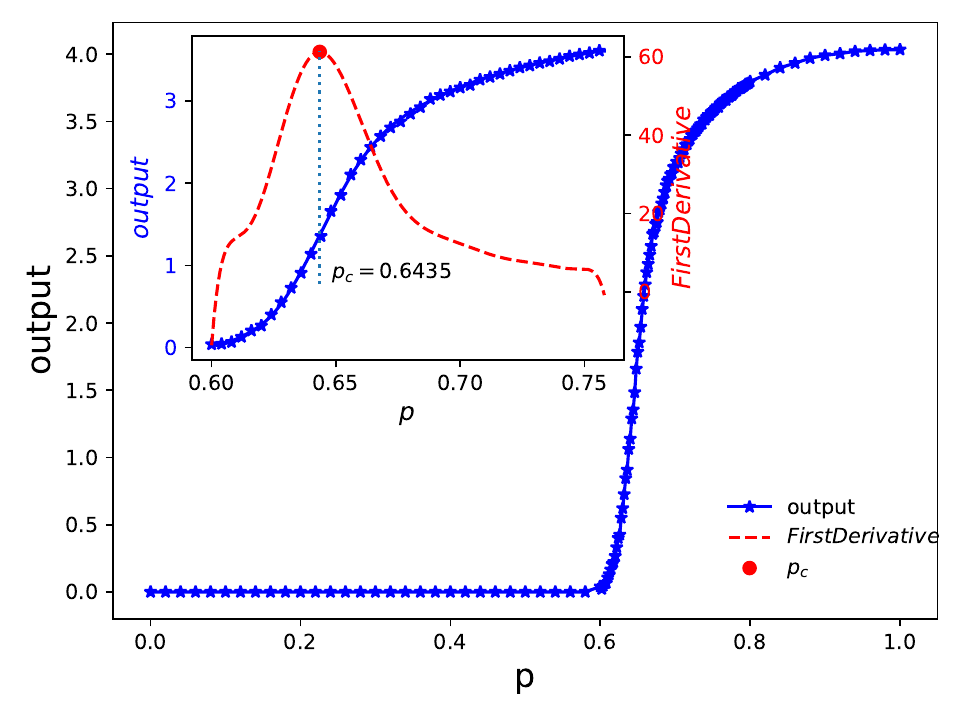} &\\
    (a) &  (b)\\
    \includegraphics[width=0.45\textwidth]{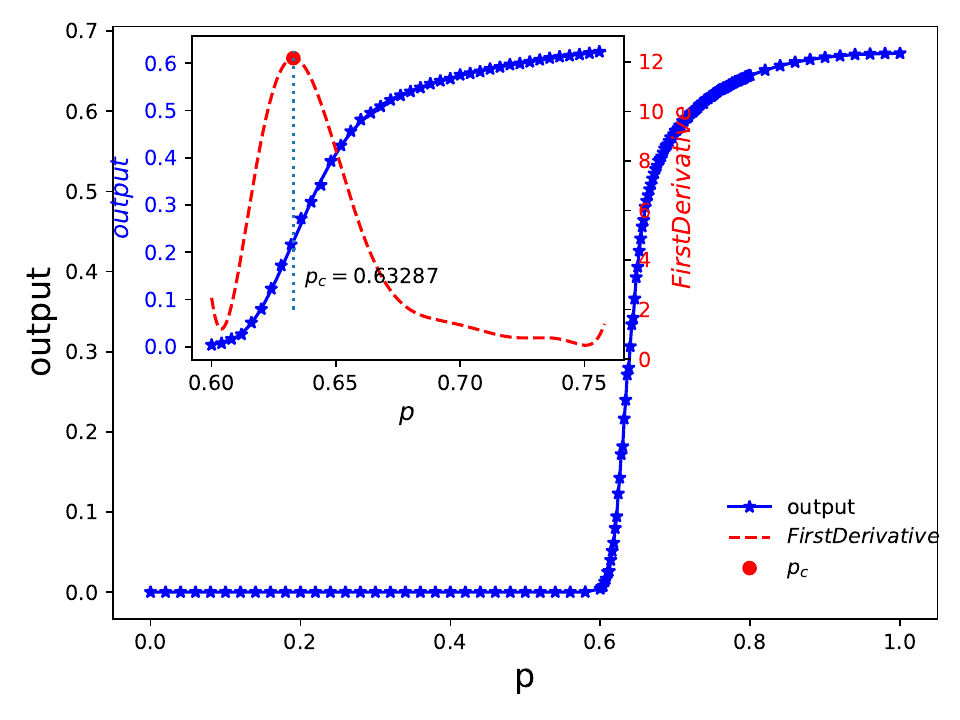} &
    \includegraphics[width=0.45\textwidth]{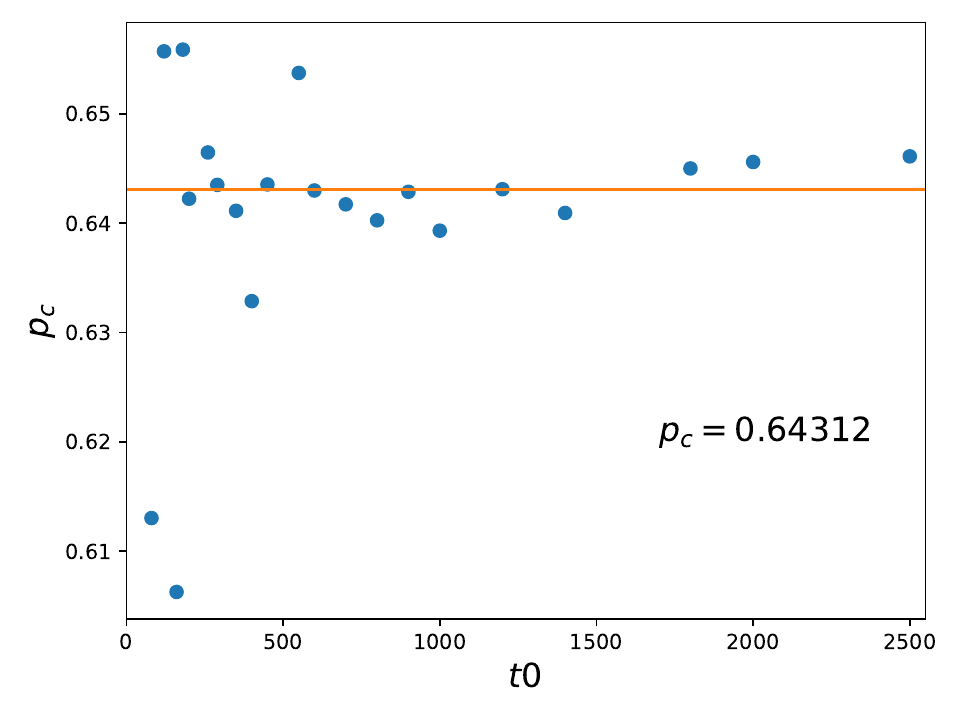} &\\
     (c)&  (d)\\
\end{tabular}
\caption{The output results of the autoencoder network for the (1+1)-dimensional DP model at different single time steps are shown. \textbf{(a)}-\textbf{(c)} display the results for $t_0=200$, $t_0=300$, and $t_0=400$, respectively. \textbf{(d)} shows the critical point results for different single time steps, with the steady-state average value ($t_0>500$) being $p_c=0.64312$.}
\label{mc_pc}
\end{figure*}


The left side of Fig~\ref{configuration} shows the configurations of the (1+1)-dimensional DP model with system size $L=64$ and evolution time from $t=0$ to $t=100$.The right side of Fig~\ref{configuration} displays the single-step configurations selected from these configurations at specific time steps. Fig~\ref{mc_ans}(a) and Fig~\ref{mc_ans}(c) show the average particle number density of the (1+1)-dimensional and (2+1)-dimensional DP models at different probabilities $p$.When the time step $t_0$ is sufficiently large, the single-step configuration can serve as a substitute for the overall configuration in identifying the critical point.

For both (1+1)-dimensional and (2+1)-dimensional DP models, single-step configurations at a specific time \( t_0 \) are used as input to the network. Monte Carlo simulations generate configurations at different probabilities \( p \), and the particle number density of the configurations is analyzed. For the one-dimensional DP model, the particle density approaches the average density over all time steps as time progresses. In the two-dimensional DP model, after a sufficient number of time steps, the particle density at each time step behaves similarly to that of the single-step configuration. The system exhibits long-range correlations, particularly near the critical point. Single-step configurations after a long enough time can represent the system’s critical behavior, which significantly reduces the training dataset size for machine learning.

Through Monte Carlo (MC) simulations, configurations are generated at different probabilities \( p \), and the average particle density is calculated. Then, the first derivative is computed by fitting a high-order polynomial, and it is observed that the first derivative reaches its maximum at the critical point \( p = p_c \). As shown in the Fig~\ref{mc_ans}(b) and Fig~\ref{mc_ans}(d), the critical points for different system sizes are calculated, and finite-size scaling yields critical points \( p_c = 0.6445 \) for the (1+1)-dimensional system and \( p_c = 0.2867 \) for the (2+1)-dimensional system.

\begin{figure*}[!thb]
\centering
    \includegraphics[width=0.98\textwidth,trim=100 50 10 50,clip]{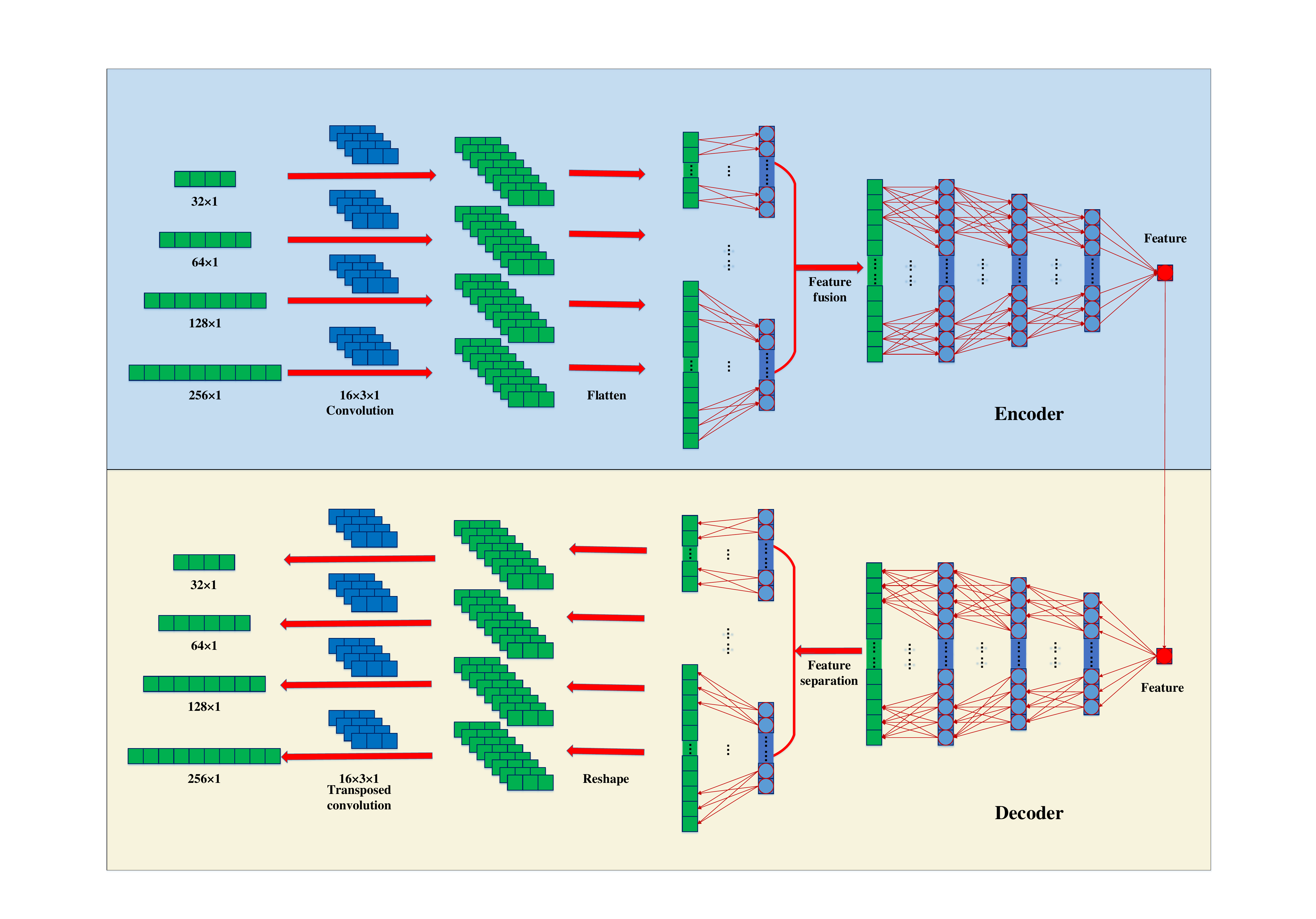} 
    \caption{Neural network schematic structure of multi-input branch autoencoder network.}
    \label{msnn}
\end{figure*}

The single-step configurations generated by Monte Carlo simulations are used as input to the autoencoder network, which is trained to extract a single latent variable representing the data. In general, this latent variable corresponds to the system's order parameter, allowing the neural network to directly identify the critical point without knowledge of the physical meaning of the model's order parameter.

For the (1+1)-dimensional DP model with system size $L=128$, the network’s output varies with probability $p$ for different single-step times.By fitting the results with a high-order polynomial, the position of the maximum value of the first derivative is identified as the critical point.
Fig~\ref{mc_pc}(a), Fig~\ref{mc_pc}(b), and Fig~\ref{mc_pc}(c) show the results for $t_0=200$, $t_0=300$, and $t_0=400$, , respectively, with minimal deviation between them. Figure 5d displays the critical points obtained for different single-step times $t_0$. When $t_0$ is small, the results show greater deviation, but as $t_0$ increases, the results become more stable. This behavior is due to the system needing sufficient time to reach a non-equilibrium steady state, which is influenced by both system size and probability. Once the steady state is reached, a single time-step configuration can adequately represent the system’s behavior across all time steps. The average critical point obtained after reaching steady state is $p_c=0.64312$, which agrees with the theoretical result~\cite{domany1981directed}.

\section{multi-input branch autoencoder network}
Due to finite-size effects, the critical point location is influenced by system size. Typically, finite-size scaling methods are used to obtain the critical point for an infinite system. This paper proposes a multi-input branch autoencoder network, and the network results are shown in the Fig~\ref{msnn}. The framework uses an autoencoder, consisting of an encoder and a decoder. The encoder takes configurations of different system sizes as input, and these inputs undergo convolution operations before passing through a fully connected layer to extract features corresponding to each system size. The features are then fused and input into multiple fully connected layers, which output a single latent variable. The decoder performs the inverse decoding operation on the latent variable output by the encoder to reconstruct the input configuration. The network simultaneously extracts shared features from configurations of different sizes, and these shared features exhibit scale invariance.

\begin{figure*}[htbp]
\centering
\begin{tabular}{cccc}   
    \includegraphics[width=0.45\textwidth]{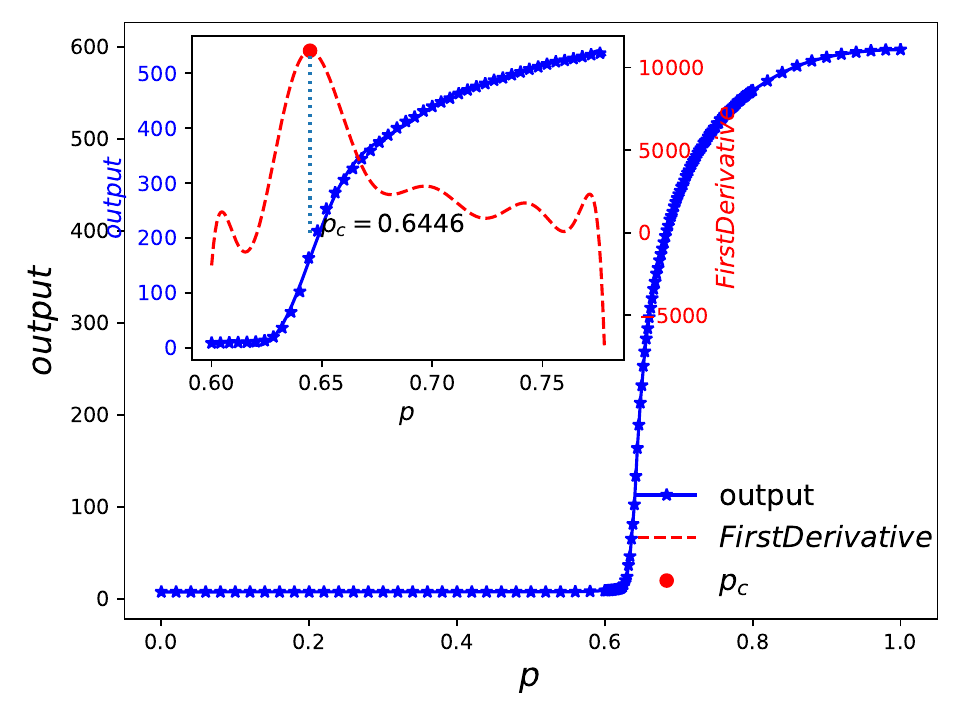} &
    \includegraphics[width=0.45\textwidth]{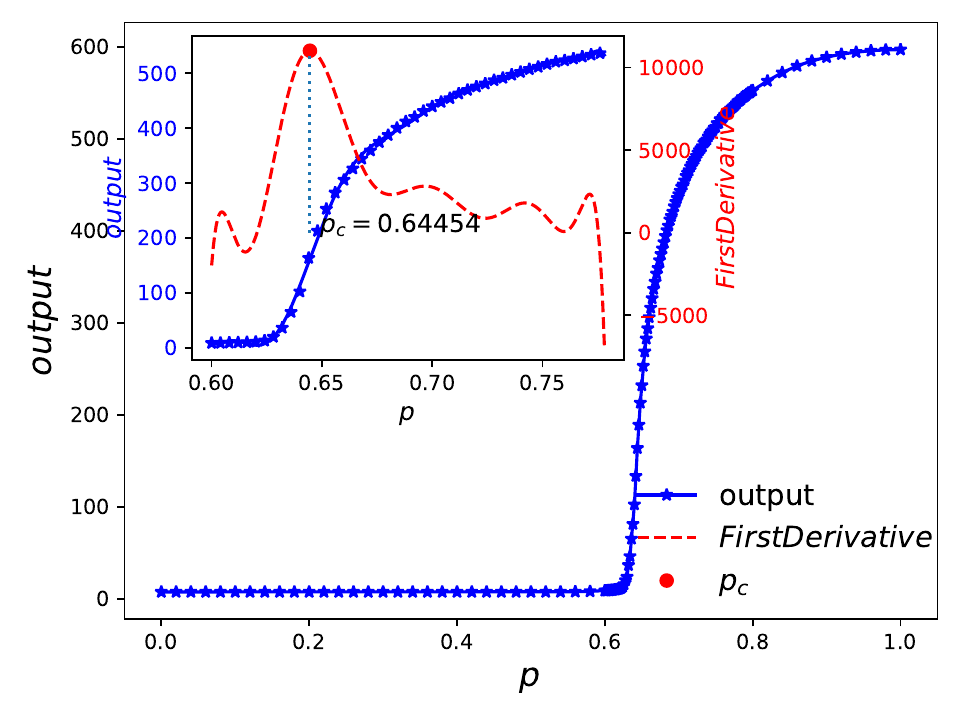} &\\
    (a) &  (b)\\
    \includegraphics[width=0.45\textwidth]{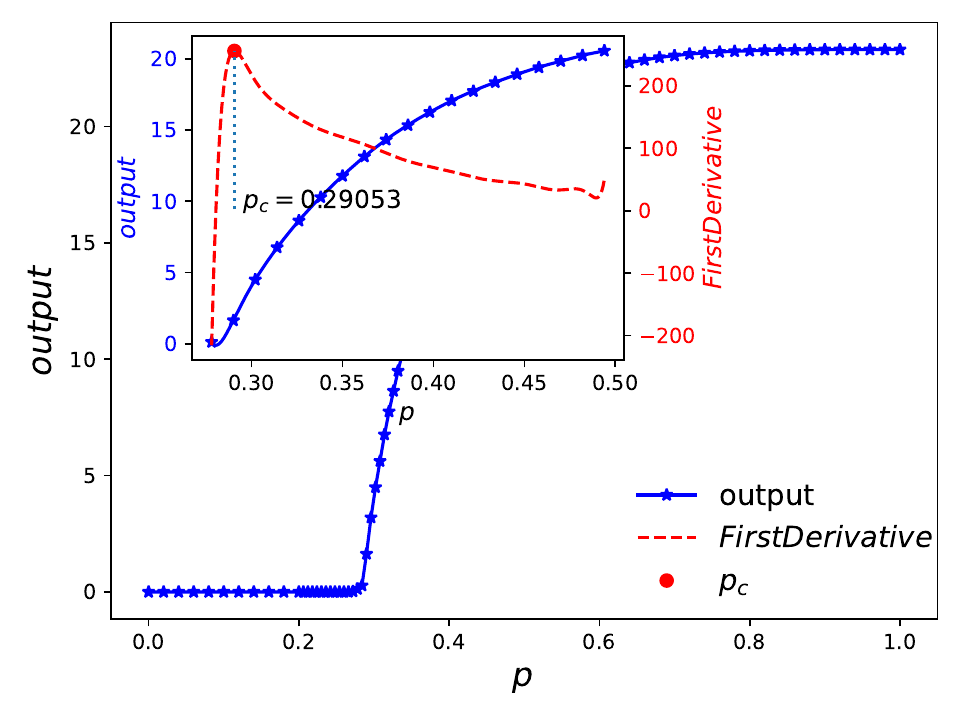} &
    \includegraphics[width=0.45\textwidth]{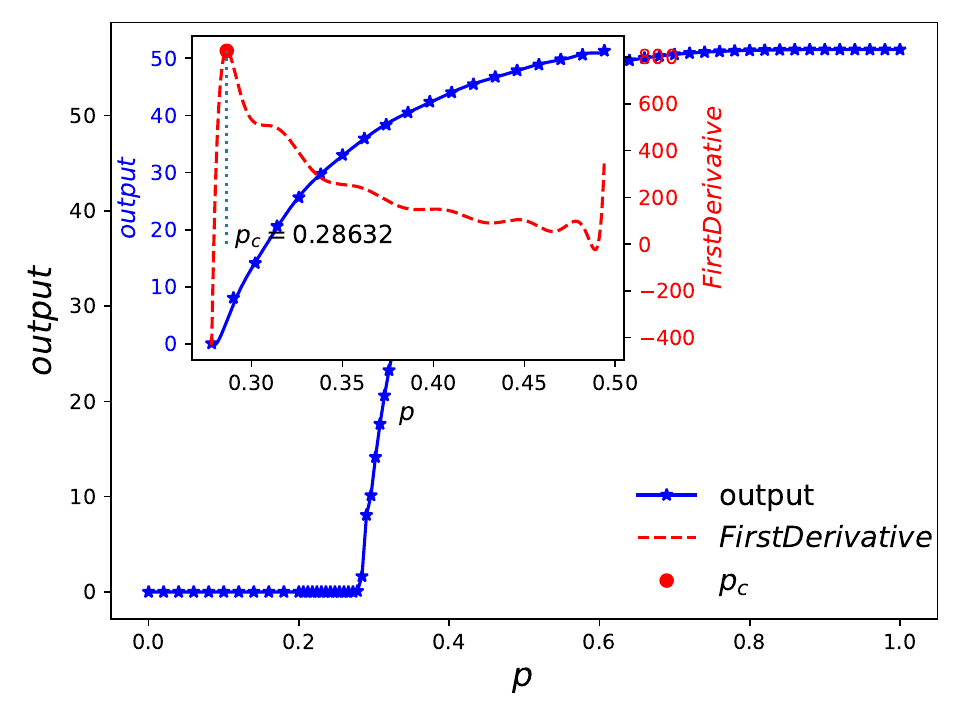} &\\
     (c)&  (d)\\
\end{tabular}
\caption{The results of the multi-input branch autoencoder network for the (1+1)-dimensional DP (top) and (2+1)-dimensional DP (bottom) models are shown. \textbf{(a)} and \textbf{(c)} display the results where the neural network extracts features from systems of different sizes. \textbf{(b)} and \textbf{(d)} show the results where the neural network captures discrete temporal information.}
\label{msnn_pc}
\end{figure*}

This model is a multi-input channel autoencoder designed to handle data from different scales and extract shared features from the data through deep learning. The network consists of two main parts: the encoder and the decoder. The encoder first processes the input data (e.g., 32×1, 64×1, 128×1, 256×1) using multiple convolutional layers to extract local features and gradually reduce the dimensionality. Each input channel’s data undergoes convolution with a \(16 \times 3\) convolutional kernel, followed by a Flatten layer to convert the output features from the convolution layers into one-dimensional vectors. Then, two fully connected layers further compress the features, and ultimately, the features from different scales are fused into a low-dimensional representation for the decoder to use.

The decoder reconstructs the compressed features from the encoder using deconvolution layers (i.e., transpose convolution), progressively rebuilding the original form of the input data. First, the decoder uses two fully connected layers to gradually expand the compressed features to the target shape, and then applies deconvolution layers to reshape the expanded features into the same dimension as the original input, completing the data reconstruction. The autoencoder uses Mean Squared Error (MSE) as the loss function to optimize the model parameters, minimizing the error between the reconstructed data and the original input. Through the fusion and compression of multi-scale data, the network learns the shared features across the data and achieves efficient data representation.

The model’s advantage in handling multi-scale data makes it particularly suitable for tasks that require integrating information from different sources or scales. By feature fusion and compressed representation, the network efficiently extracts latent patterns and achieves a good balance between reconstruction accuracy and training efficiency.

This study employs Monte Carlo simulations to generate configurations of the (1+1)-dimensional and (2+1)-dimensional directed percolation models at various probabilities. For the (1+1)-dimensional model, four different system sizes (\( L = 32, 64, 128, \) and \( 256 \)) are used as input. Each input corresponds to a single-step configuration taken after a sufficiently long evolution time, such as \( t_0 = 200 \) for \( L = 32 \).

The data is fed into the network, and the average output at different probabilities is computed. By fitting the results with a high-order polynomial, the location of the maximum first derivative is identified as the critical point. The obtained critical probabilities are \( p_c = 0.6446 \) for the (1+1)-dimensional model and \( p_c = 0.29047 \) for the (2+1)-dimensional model shown in Fig~\ref{msnn_pc}(a) and Fig~\ref{msnn_pc}(c). By extracting shared features across different system sizes, the network outputs a single latent variable, allowing the identification of the system’s critical point. This critical point exhibits scale invariance, aligning with the results of finite-size scaling analysis.

By inputting configurations of different system sizes, the network can extract shared features from them. Similarly, we can input configurations at different times to capture temporal information. For instance, in the (1+1)-dimensional DP model, we can input multiple single-step configurations for \( L = 32 \) at different times (e.g., \( t_1 = 100, t_2 = 200, t_3 = 300, t_4 = 400\). The process follows the same approach as before, where shared features across different time steps are extracted to determine the critical point. The results are shown in Fig~\ref{msnn_pc}(b) and Fig~\ref{msnn_pc}(d), and they are consistent with those from other methods.


The network extracts a single latent variable that represents the shared features across different time steps. These latent variables capture dynamic information over time, making the model capable of handling discrete dynamic information. Thus, the proposed network can also be applied to analyze dynamic information of non-equilibrium phase transitions, helping identify the location of the critical point. If data from a sufficiently large number of time steps is available, the model can also handle continuous dynamic information.

\section{Conclusion}

In this paper, we propose a multi-input branch autoencoder network for extracting both spatial and temporal features from systems undergoing non-equilibrium phase transitions, specifically within the context of directed percolation models. By inputting configurations of different system sizes or configurations at different time steps, the network is able to capture shared latent variables that reflect both spatial and dynamic information.

For the (1+1)-dimensional and (2+1)-dimensional DP models, we demonstrated that the network can accurately identify the critical point by extracting relevant features at different system sizes and times. The critical points for the two models were obtained as \( p_c = 0.6446 \) and \( p_c = 0.29047 \), respectively, consistent with theoretical results. The use of single-step configurations at different time points allows for the extraction of discrete dynamic information, enabling the identification of the critical point through high-order polynomial fitting and the first derivative maximum.

Moreover, the proposed network offers flexibility in that it can handle both discrete and continuous dynamic information. When sufficient time step data is available, the network can capture continuous dynamic behavior, further enhancing its application to non-equilibrium phase transitions.

The primary advantage of the proposed multi-input branch autoencoder network lies in its ability to handle data across different scales and times, providing a comprehensive framework for studying non-equilibrium systems. This approach allows for the identification of critical points with high accuracy, using both spatial and temporal features, and can be generalized to study a range of dynamic phase transitions.

By incorporating configurations from multiple time steps, the network can learn the system's time-dependent behavior, offering a powerful tool for analyzing non-equilibrium dynamics. Furthermore, the network’s ability to handle different system sizes makes it particularly suitable for systems where scaling effects play a significant role.

Future work could explore the use of this approach in more complex non-equilibrium systems, potentially incorporating other dynamic models or testing the network on experimental data. Additionally, exploring the network's performance on larger systems or in higher-dimensional models could provide further insights into its scalability and robustness.

\section{Acknowledgements}
This work was supported in part by Yunnan Fundamental Research Projects (Grant 202401AU070035), Research Fund of Baoshan University(BSKY202305), and the 111 Project 2.0, with Grant No. BP0820038.

\section{Data sets}
The detailed algorithms of how to generate raw data and implement machine learning are shown in the GitHub link {https://github.com/ChuckShen/perc-snn}.

\bibliographystyle{apsrev4-2}
\bibliography{ref}

\end{document}